\documentclass[twocolumn]{aastex631}
\usepackage{amsmath}

\newcommand{\sigmanu}{$\Sigma m_{\nu}$}
\newcommand{\summnu}{\Sigma m_{\nu}}

\newcommand{\lya}{Ly$\alpha$}
\newcommand{\lyaf}{Ly$\alpha$ forest}
\newcommand{\pd}{$P_\mathrm{1D}$}
\newcommand{\Mpc}{\ \text{Mpc}}
\newcommand{\iMpc}{\ \text{Mpc}^{-1}}
\newcommand{\kms}{{\rm km~s^{-1}}}
\newcommand{\ikms}{{\rm s~km^{-1}}}

\newcommand{\thetamc}{\theta_\mathrm{MC}}

\begin{document}

\title{Compressing the cosmological information in one-dimensional correlations
of the Lyman-$\alpha$ forest}
\email{c.pedersen@nyu.edu, afont@ifae.es, gnedin@fnal.gov}

\author[0000-0002-4315-9295]{Christian Pedersen}
\affiliation{Center for Cosmology and Particle Physics, Department of Physics, New York University, New York, NY 10003, USA}
\affiliation{Center for Computational Astrophysics, Flatiron Institute, New York, NY 10010, USA}
\affiliation{Department of Physics and Astronomy, University College London, London WC1E 6BT, United Kingdom}
\affiliation{Fermi National Accelerator Laboratory; Batavia, IL 60510, USA}

\author[0000-0002-3033-7312]{Andreu Font-Ribera}
\affiliation{Institut de F\'isica d’Altes Energies (IFAE), The Barcelona Institute of Science and Technology, 08193 Bellaterra (Barcelona), Spain}
\affiliation{Department of Physics and Astronomy, University College London, London WC1E 6BT, United Kingdom}

\author[0000-0001-5925-4580]{Nickolay Y. Gnedin}
\affiliation{Fermi National Accelerator Laboratory; Batavia, IL 60510, USA}
\affiliation{Kavli Institute for Cosmological Physics; The University of Chicago; Chicago, IL 60637 USA}
\affiliation{Department of Astronomy \& Astrophysics; The University of Chicago; Chicago, IL 60637 USA}

\begin{abstract}Observations of the Lyman-$\alpha$ (\lya) forest from spectroscopic surveys such as BOSS/eBOSS, or the ongoing DESI, offer a unique window to study the growth of structure on megaparsec scales. 
Interpretation of these measurements is a complicated task, requiring hydrodynamical simulations to model and marginalise over the thermal and ionisation state of the intergalactic medium.
This complexity has limited the use of \lya\ clustering measurements in joint cosmological analyses.
In this work we show that the cosmological information content of the 1D power spectrum (\pd) of the \lyaf\ can be compressed into a simple two-parameter likelihood without any significant loss of constraining power.
We simulate \pd\ measurements from DESI using hydrodynamical simulations and show that the compressed likelihood is model independent and lossless, recovering unbiased results even in the presence of massive neutrinos or running of the primordial power spectrum.\end{abstract}


\section{Introduction} 
\label{sec:introduction}

The tightest constraints on cosmological parameters are obtained from the joint analysis of complementary probes, with different sensitivity to cosmological parameters.
A common approach is to combine observations of the cosmic microwave background (CMB) with late-time probes of large-scale structure (LSS), such as galaxy clustering or weak lensing \citep{Planck2018,eBOSS2021,DES-Y3}.
An alternative probe of LSS is the Lyman-$\alpha$ (\lya) forest, a series of absorption features in the spectra of $z>2$ quasars, caused by intervening neutral hydrogen along the line-of-sight.

Cosmological analysis of the \lyaf\ is driven by large spectroscopic surveys, such as the Baryon Oscillation Spectroscopic Survey (BOSS, \citep{BOSS2013}) and its extension eBOSS \citep{eBOSS2016}, which between 2009 and 2019 observed $\sim 200,000$ \lyaf\ quasars.
In 2021, the Dark Energy Spectroscopic Instrument (DESI) \citep{DESI2016} started a five-year program to survey a third of the sky and obtain spectra of $\sim 800,000$ \lyaf\ quasars. 
The main goal of these quasar surveys is to measure the 3D correlations in the \lyaf\ and to provide accurate measurements of the Baryon Acoustic Oscillations (BAO) feature to study the expansion of the universe \citep{eBOSS2020}.
The same dataset, however, can be used to measure correlations along the line of sight, known as the 1D flux power spectrum (\pd), a unique window to study the clustering of matter on megaparsec scales \citep{Chabanier2019b}.

Cosmological analyses of the \pd\ are particularly powerful in combination with CMB measurements due to the large ``lever arm'' between the two measurements, and these joint analyses have historically provided some of the tightest constraints on the sum of the neutrino masses, and on the shape of the primordial power spectrum of density fluctuations \citep{Phillips2001,Verde2003,Spergel2003,Viel2004b,Seljak2005,Seljak2006,Bird2011,PD2015,PD2015b,PD2020}.

Massive neutrinos are known to affect the growth of structure by suppressing the late-time clustering of matter on scales smaller than their free-streaming length \citep{Julien06}. 
The \pd\ alone is unable to constrain neutrino masses due to parameter degeneracies \citep{Pedersen2020}, but when combined with the early-time, large-scales measurements from the CMB one can break these degeneracies.
In the next few years, and in combination with CMB measurements, several LSS probes will be able to detect the impact of massive neutrinos, even if the sum of the masses is near the minimum of $\summnu=0.06$ eV allowed by oscillation experiments \citep{2014JCAP...05..023F}. 

At the same time, inflationary models generically predict that the primordial power spectrum of fluctuations should have small deviations from a power law, often parameterised as a \textit{running} of the spectral index. 
Due to the wide lever arm between the large scale fluctuations probed by \textit{Planck} and the small scales accessed by the \pd, the \lyaf\ is one of the most promising avenues towards tightening the constraints on inflationary models which produce a measurable running of the spectral index \citep{2014JCAP...05..023F}.

Unfortunately for cosmologists, the statistical properties of the \lyaf\ also depend on the thermal and ionisation history of the intergalactic medium (IGM) \citep{McQuinn2016} 
\footnote{This also makes the \lya\ forest, specially at $z>5$, a key probe of reionisation, but we do not discuss this in this work.}.
This has two consequences that complicate \pd\ analyses. 
First, it means that we need to run expensive hydrodynamical simulations in order to make accurate predictions for a given model.
Second, it means that we need to add multiple nuisance parameters in our cosmological inference, and to carefully marginalise over them to obtain robust cosmological constraints.

In the last few years, several groups have attempted to tackle the first
problem, introducing new tools to \textit{emulate} \pd\ for parameters
that are not covered by the relatively small suite of simulations available
\citep{Walther2018,Bird2019,Rogers2019,Takhtaganov2021,Rogers2021a,Pedersen2021}.
In this publication, we will use the \texttt{LaCE}
\footnote{https://github.com/igmhub/LaCE} emulator presented in
\citet{Pedersen2021}, and focus on the second problem: the high-dimensionality
of the parameter space sampled, and the attractive possibility of dramatically
reducing the dimensionality of the \pd\ likelihood into a small number
of parameters describing the linear matter power spectrum,
without introducing biases or losing relevant information.

The idea of compressing the \pd\ likelihood into a handful of parameters describing the linear power spectrum is not new. 
Indeed, the first cosmological studies of the \lya\ forest focused on recovering the matter power spectrum \citep{Croft1998,McDonald2000,Croft2002,Gnedin2002},
and the two-parameter (amplitude and slope) parameterisation we
focus on in this work was already used 20 years
ago \citep{McDonald2000}.
However, most recent \pd\ analyses from BOSS and eBOSS surveys have only presented their results in terms of direct fits to the traditional $\Lambda$CDM parameters \citep{Borde2014,PD2015,PD2015b,PD2020}, with strong dependence on the priors chosen.
This has made it difficult for other groups to include these powerful results into combined cosmological analyses. If the \lya\ forest constraints from \pd\ could be accurately and losslessly represented by just the amplitude and a local slope at a conveniently chosen pivot scale, it would significantly simplify the combination of \lya\ forest measurements with other cosmological probes.

Motivated by the latest \pd\ measurements from eBOSS, the start of the DESI survey, and the recent developments in emulation techniques, in this publication we review the compression of the \pd\ likelihood.
Note that similar discussions are also happening in the context of analysis of the galaxy power spectrum, in particular regarding the information content in measurements of redshift space distortions 
\citep{Hamann2010,Ivanov2020,Amico2020,Brieden2021}.

We will start in Section \ref{sec:methods} with a description of the simulated data, a summary of the emulator used, and the parameterisation of the likelihood.
In Section \ref{sec:direct} we present cosmological constraints from simulated \pd\ data, and discuss the impact of priors and model depencency of the results.
In Section \ref{sec:cmb} we present joint fits when combining the \pd\ with an approximated CMB likelihood, and show that the \pd\ likelihood can be efficiently compressed into two parameters without any loss of information.
Finally in Section \ref{sec:discussion} we discuss our findings.
\section{Methodology}
\label{sec:methods}

We discuss here the \lyaf\ \pd\ likelihood, including an overview of the
emulator used to make theoretical predictions (based on \citep{Pedersen2021}),
a description of the mock dataset, and a discussion of the parameterisation
of the likelihood.

\subsection{Simulations}
\label{ss:simulations}
We begin by describing the simulations used in the analysis, which
fall into two categories. First, a set of \textit{training }
simulations are used to construct the emulator. Second, a small number of
\textit{test} simulations are run, to represent mock \pd\ measurements in
a variety of different cosmologies, and are used to test and validate our
analysis pipeline.
Both the training and most of the test simulations were presented in 
\citet{Pedersen2021}, where we also described and tested the emulation framework.
Here we give an overview of the simulations, and refer the reader to 
\citet{Pedersen2021} for a more detailed description.

The simulations were run in \texttt{MP-Gadget}
\footnote{\url{https://github.com/MP-Gadget/MP-Gadget}.} \citep{mp-gadget},
a \texttt{TreeSPH} code based on \texttt{Gadget-2} \citep{Gadget2}.
All simulation boxes had a size of $L=67.5$ Mpc and $768^3$ gas and cold dark
matter (CDM) particles.
The initial conditions were generated with \texttt{MP-GenIC} at $z=99$,
with Fourier modes that had random phases but fixed initial amplitudes
\citep{Angulo2016,Anderson2019,FVN2018,Pedersen2021}.
In order to further reduce cosmic variance, for each model (both in the
training and test sets) we ran a pair of simulations with inverted phases, 
and each quantity estimated from the simulations is taken
as the average of the pair.

We output 11 snapshots, equally spaced in redshift between $z=2$ and $z=4.5$. 
To produce mock \lyaf\ spectra from each snapshot, we use \texttt{fake\_spectra}
\footnote{\url{https://github.com/sbird/fake_spectra}.} \citep{fakespec}
to calculate a 2D grid of $500^2$ transmission skewers from each snapshot,
with a line-of-sight resolution of $0.05$ Mpc.
The cosmological and astrophysical parameters used in both the training and
test simulations are listed in Table \ref{tab:sims}.

\begin{table}[]
  \centering
  \begin{tabular}{|l|l|l|l|l|}
  \hline
    & Training set & \textit{Central} & \textit{Neutrino} & \textit{Running} \\
  \hline
  $A_s(\times10^{-9})$ & $[1.35\mbox{--}2.71]$ & 2.006 & 2.251 & 2.114 \\
  $n_s$                & $[0.92\mbox{--}1.02]$ & 0.9676 & 0.9676 & 0.9280 \\
  $\alpha_s$           & 0.0 & 0.0 & 0.0 & 0.015 \\
  $\Omega_m$           & 0.316 & 0.316 & 0.324 & 0.316 \\
  $\summnu$  (eV)      & 0.0 & 0.0 & 0.3 &  0.0 \\ \hline
  $\Delta^2_p(z=3)$    & $[0.25$--$0.45]$ & \multicolumn{3}{c|}{0.35} \\
  $n_p(z=3)$           & $-[2.35\mbox{--}2.25]$ & \multicolumn{3}{c|}{-2.30} \\
  \hline
  $z_\mathrm{rei}$     & $[5.5\mbox{--}15]$ & \multicolumn{3}{c|}{10.5} \\
  $H_A$                & $[0.5\mbox{--}1.5]$ & \multicolumn{3}{c|}{1.0} \\
  $H_S$                & $[0.5\mbox{--}1.5]$ & \multicolumn{3}{c|}{1.0} \\
  \hline
  \end{tabular}
  \caption{Cosmological and astrophysical parameters for the training
  and test simulations. 
  The limits of the Latin hypercube for the training simulations are shown
  in the left column, where only the primordial power spectrum and
  astrophysical parameters are varied.
  The primordial parameters $A_s$ and $n_s$ here are defined at the CMB pivot
  scale of $k=0.05\:\mathrm{Mpc}^{-1}$.
  The \textit{Central}, \textit{Neutrino} and \textit{Running} simulations are
  constructed such that they have the same small scale linear matter power
  spectrum ($\Delta^2_p$ and $n_p$) at $z=3$. 
  For all simulations, we fix $\omega_c=0.12$, $\omega_b=0.022$ and $h=0.67$.}
  \label{tab:sims}
\end{table}

\subsection{Emulator}
\label{ss:emulator}

We provide a brief overview of the emulator parameters and framework.
We use the \texttt{LaCE}\footnote{\url{https://github.com/igmhub/LaCE}.} 
framework presented in \citet{Pedersen2021}, and refer the reader to this 
reference for a more complete description.
\texttt{LaCE} uses a Gaussian Process emulator\footnote{We use the Python implementation \texttt{GPy} \citep{gpy2014}.}
to predict \pd\ as a function of six parameters:
the dimensionless amplitude ($\Delta^2_p$) and slope ($n_p$) of the linear
power spectrum around a pivot scale of $k_p=0.7 \iMpc$;
the mean transmitted flux fraction (or mean flux, $\bar F$);
a thermal broadening scale defined in comoving units ($\sigma_T^{\rm com}$),
set by the temperature of the gas at mean density;
the slope of the temperature-density relation ($\gamma$);
the filtering length in inverse comoving units ($k_F^{\rm com}$),
a proxy for gas pressure. 
Note that whilst the \pd\ is naturally observed in velocity units,
the above quantities are all defined in comoving units in the emulator. 
The motivation for this is so that the simulated \pd\ is estimated at a
fixed set of wavenumbers for snapshots at all redshifts.

We train the emulator using 30 pairs of simulations described in Table
\ref{tab:sims}.
These simulations explore different thermal and reionisation histories
by varying $z_\mathrm{rei}$, $H_A$, and $H_s$, which control
the redshift of reionisation and the heating rates of the gas as rescalings around a fiducial model from \cite{HM2012} (see
\citet{Pedersen2021} for more detail).
The simulations have
different amplitudes and slopes of the primordial power spectrum ($A_s$, $n_s$),
but have the same value for the physical densities of CDM
($\omega_c=\Omega_c h^2$) and baryons ($\omega_b=\Omega_b h^2$),
the same value of $H_0$, and do not include massive neutrinos.
All 11 snapshots from all 30 models are used simultaneously, for a total of
330 points in the training sample.

In the implementation of the emulator presented in \citet{Pedersen2021},
the \lya\ \pd\ was emulated directly on a grid of comoving wavenumbers.
Because of the limited box size of our simulations, the \pd\ measurements on
large scales are affected by cosmic variance.
Given that each simulation was run with the same random seed,
there are random noise spikes in the power spectrum that align at the same
comoving wavenumbers in each simulation used to train the emulator.
When it came to testing the pipeline on simulated mock data with a different
background evolution, we found that these noise spikes are interpreted
by the emulator as sharp features in the power spectrum, artificially
enhancing the emulator sensitivity to changes in cosmology.
This is due to the fact that the likelihood evaluation is performed in
velocity units, which require a conversion from comoving units using $H(z)$.
The pipeline would therefore try and find the $H(z)$ which would align the
noise spikes in the observed data with the noise spikes in the training set,
with two consequences.
When running on mock simulations with the same training seed, the pipeline is
artificially sensitive to the conversion between comoving and velocity units,
and therefore $H(z)$, due to the presence of these sharp features.
When running on mock simulations with a different random seed, the pipeline
struggles to return the correct cosmology, as the likelihood maximisation
is dominated by the incentive to align the different noise features.

In order to remove residual noise in the measurements of \pd, we fit a 4th
order polynomial 
\footnote{This setting performed better than 3rd or 5th
order polynomials, but we did not explore other functions.}
to the logarithm of \pd\ as a function of the logarithm of wavenumber, 
using scales $k_\parallel < 8 \iMpc$:
\begin{equation}
 \log P_{\rm 1D}(k_\parallel) = \sum_{n=0}^4 c_n 
        ~ \big(\log k_\parallel \big)^n ~.
 \label{eq:polyfit}
\end{equation}
Instead of predicting directly \pd\, the emulator now predicts the five
coefficients $c_n$ of this polynomial, that can later be used to predict
\pd\ on all scales. The use of fitting functions, such as polynomials
or principle component analysis is a standard practice to reduce noise in
emulators, such as in the ``Coyote" emulator \citep{Lawrence2010} as well as in
early analyses of \pd\ \citep{McDonald2005}.

The variance on the emulated coefficients ($\sigma_{c_n}^2$) can be
used to obtain an estimate for the variance of the emulated \pd:
\begin{equation}
 \Big(\frac{\sigma_{P_{\rm 1D}}}{P_{\rm 1D}}\Big)^2  
        = \sum_{n=0}^4 \sigma^2_{c_n} \big( \log k_\parallel \big)^{2n} ~.
 \label{eq:emu_cov_polyfit}
\end{equation}
We discuss the impact of cosmic variance on emulator predictions in Appendix
\ref{app:seed}.

\subsection{Mock data}
\label{ss:mocks}

In order to test our analysis pipeline, we have generated three synthetic
datasets (or \textit{mocks}) for models that are not included in the
training set of the emulator:

\begin{itemize}
  \item \textit{Central} simulation: this is the simplest case, a simulation 
without massive neutrinos or running, with the same background expansion as was used
in all training simulations, and a primordial power ($A_s$,$n_s$) corresponding
to the centre of the Latin Hypercube used to setup the training set.

  \item \textit{Neutrino} simulation: a simulation with $\summnu=0.3$ eV, where
the cosmological constant ($\Lambda$) has been lowered to compensate the
increase in the total matter density.
The amplitude of the primordial power is also $\sim 10$\% larger to compensate
the suppression of power caused by massive neutrinos.
In \citet{Pedersen2021} we used this simulation to show that we could recover
unbiased predictions in cosmologies with massive neutrinos, even when the
emulator was trained exclusively with simulations with massless neutrinos. 

  \item \textit{Running} simulation: a simulation with the same cosmology
than the \textit{Central} simulation, except that its primordial power
spectrum has a non-zero running of $\alpha_s=0.015$.
The other parameters describing the primordial power ($A_s$, $n_s$) have
been modified to compensate the change in running and have the same linear
power around the pivot scale used in the emulator ($k_p=0.7\iMpc$, see Table
\ref{tab:sims}).

\end{itemize}

We start by running a pair of simulations (with inverted phases) for each of
the three test models.
From each of their 11 snapshots we measure \pd, in comoving units, and fit a 
4th order polynomial as described in Section \ref{ss:emulator} above.
In order to roughly simulate the statistical power of DESI, we use a rescaled
version of the SDSS DR14 covariance matrix of \citet{Chabanier2019}, where all
elements are divided by 5 to approximately take into account the difference in
the number of spectra between SDSS DR14 and DESI
\footnote{A more detailed forecast should also take into account the
differences in pixelisation, spectral resolution and signal to noise, but we
leave this for future work.}.
As is common in \pd\ measurements, the band powers presented in
\citet{Chabanier2019} are defined in velocity units.
At each redshift we compute $H(z)/(1+z)$ using the simulation cosmology to
translate these into wavenumbers in comoving units.

\subsection{Likelihood}
\label{ss:likelihood}

We use a Gaussian likelihood, naturally decomposed into 11 independent
sub-likelihoods, one for each snapshot (redshift bin).
The covariance matrix is the sum of the data covariance and an extra term
describing the uncertainty in the emulator predictions, computed with Equation
\ref{eq:emu_cov_polyfit}.
The typical emulator uncertainty is smaller than 1\% for models near the
centre of our training set, and it only has a minor impact on likelihood
evaluations around the best-fit values of our analyses.
However, it can be larger than 10\% when evaluating the likelihood near
the convex hull of our training sample.

Different sub-sections in Section \ref{sec:direct} and Section \ref{sec:cmb} use a
different number of free cosmological parameters, including:
the amplitude ($A_s$), slope ($n_s$) and running ($\alpha_s$) of the primordial
power spectrum at the usual CMB pivot scale of $k_s=0.05 \iMpc$;
the physical densities of baryons ($\omega_b=\Omega_b h^2$) and of CDM
($\omega_c=\Omega_c h^2$);
the sum of the neutrino masses (\sigmanu);
the Hubble parameter $H_0$;
the angular acoustic scale of the CMB ($\thetamc$).

We use four functions to describe thermal and ionisation history of the IGM:
the effective optical depth as a function of redshift $\tau(z)=-\log \bar F(z)$,
the thermal broadening scale (in $\kms$) at mean densities
$\sigma_T^{\rm vel}(z)$,
the slope of the temperature-density relation $\gamma(z)$
and the filtering / pressure scale $k_F^{\rm vel}(z)$ (in $\ikms$).
Following \citet{Pedersen2021}, we measure each of these functions from the
\textit{Central} simulation, and use two parameters, $\alpha_X$ and $\beta_X$, to describe a power law rescaling for each of the four functions.
Here the subscript $X$ refers to one of the four IGM parameters.
For instance, the thermal broadening scale, $\sigma_T^{\rm vel}(z)$ is parameterised as
\begin{equation}    
  \ln \sigma_T^{\rm vel}(z) = 
        \ln \sigma_T^{\rm vel}(z) \big \rvert_{\rm cen} 
        + a_{\sigma_T} + b_{\sigma_T} \ln \frac{1+z}{1+3} ~,
  \label{eq:sigT}
\end{equation}
where $\sigma_T^{\rm vel}(z) \big \rvert_{\rm cen}$ is the thermal
broadening scale in the \textit{Central} simulation. 
Therefore we use a total of 8 nuisance parameters related to IGM physics.

There is no guarantee that this simple parameterisation is accurate enough
to do an analysis on real data, but it should be flexible enough to test the
compression of the likelihood in a realistic setting.
As described in Table \ref{tab:priors}, we use combined priors: 
each parameter is allowed to vary within a given range
of values (top hat prior) and an additional weak Gaussian prior is
applied to all parameters; the actual prior is a product of the two.

\begin{table}[]
  \centering
  \begin{tabular}{|l|l|l|}
  \hline
  Parameter & Range allowed & Gaussian prior \\
  \hline
  $A_s(\times10^{-9})$ & $[1.0~\mbox{--}~3.2]$ & $\mathcal{N}(2.1,1.1)$ \\
  $n_s$             & $[0.89~\mbox{--}~1.05]$ & $\mathcal{N}(0.965,0.08)$ \\
  $\alpha_s$        & $[-0.8~\mbox{--}~0.8]$ & $\mathcal{N}(0.0,0.8)$ \\
  $\omega_b$        & $[0.018~\mbox{--}~0.026]$ & $\mathcal{N}(0.022,0.004)$ \\
  $\omega_c$        & $[0.10~\mbox{--}~0.14]$ & $\mathcal{N}(0.12,0.02)$ \\
  $\summnu$ (eV)    & $[0.0~\mbox{--}~1.0]$ & $\mathcal{N}(0.0,0.5)$ \\
  $H_0$             & $[50~\mbox{--}~100]$ & $\mathcal{N}(67.0,25.0)$ \\
  $\thetamc(\times10^{-3})$ & $[9.9~\mbox{--}~10.9]$ & $\mathcal{N}(10.4,0.5)$\\
  \hline
  $a_\tau$      & $[-0.1~\mbox{--}~0.1]$ & $\mathcal{N}(0.0,0.05)$ \\
  $b_\tau$      & $[-0.2~\mbox{--}~0.2]$ & $\mathcal{N}(0.0,0.1)$ \\
  $a_{\sigma_T}$  & $[-0.4~\mbox{--}~0.4]$ & $\mathcal{N}(0.0,0.2)$ \\
  $b_{\sigma_T}$  & $[-0.4~\mbox{--}~0.4]$ & $\mathcal{N}(0.0,0.2)$ \\
  $a_\gamma$    & $[-0.2~\mbox{--}~0.2]$ & $\mathcal{N}(0.0,0.1)$ \\
  $b_\gamma$    & $[-0.4~\mbox{--}~0.4]$ & $\mathcal{N}(0.0,0.2)$ \\
  $a_{k_F}$       & $[-0.2~\mbox{--}~0.2]$ & $\mathcal{N}(0.0,0.1)$ \\
  $b_{k_F}$       & $[-0.4~\mbox{--}~0.4]$ & $\mathcal{N}(0.0,0.2)$ \\
  \hline
  \end{tabular}
  \caption{Priors used for the cosmological parameters (top), and for the
  nuisance parameters describing the thermal and ionisation history of the
  IGM (bottom).
  All parameters have a limited range of values allowed, and a Gaussian prior.}
  \label{tab:priors}
\end{table}

In the next sections we will discuss constraints on two derived parameters
that are able to capture most of the cosmological information in \pd:
the (dimensionless) amplitude and slope of the linear power spectrum at
a pivot point $k_\star=0.009 \, \ikms$ and redshift $z_\star=3$
\footnote{This pivot scale was found in \citet{McDonald2005} to be optimal for
their dataset, but it might be sub-optimal for other surveys.}:
\begin{equation}
  \Delta^2_\star=\frac{k_\star^3 P_L(k_\star,z_\star)}{2\pi^2} 
  \label{eq:linP1}~,
\end{equation}
\begin{equation}
  n_\star=\frac{\mathrm{d ln} P_L(k,z)}{\mathrm{d ln}k}
      \bigg|_{k_\star, \, z_\star}
  \label{eq:linP2}
\end{equation}
where $P_L(k,z)$ is the linear power spectrum \textit{in velocity units}.
It is important to highlight that these parameters are defined in velocity
units, since \pd\ measurements are also presented in velocity units and 
parameters defined in comoving units would be model dependent.

Let us finish this section by summarising the steps needed to make a
likelihood evaluation:

\begin{itemize}
  \item Given a set of cosmological parameters, we use the Boltzman solver
    \texttt{CAMB} \citep{camb} to make predictions for $P_L(z,k)$ and $H(z)$
    at all redshifts and scales.

  \item For each redshift $z_i$ in our mock \pd\ measurement, we compute the
    value of the amplitude ($\Delta^2_p$) and slope ($n_p$) of the linear
    power, $P_L(z_i,k)$, around the pivot point $k_p = 0.7 \iMpc$.
    These are two of the six parameters that will be passed to the emulator
    to get a prediction of \pd\ at $z_i$. 

  \item The other four parameters ($\bar F$, $\sigma_T^{\rm com}$, $\gamma$,
    $k_F^{\rm com}$) are computed from the eight nuisance parameters and the
    four IGM-related functions measured from the \textit{Central} simulation.
    For instance, we use Equation \ref{eq:sigT} to compute the thermal broadening
    scale ($\sigma_T^{\rm vel}$) in velocity units at redshift $z_i$,
    and the comoving scale passed to the emulator is 
    $\sigma_T^{\rm com} = \sigma_T^{\rm vel} (1+z_i) / H(z_i)$.

  \item For each redshift, we ask the emulator to predict the \pd\
    corresponding to the six emulator parameters computed above. 
    The emulator prediction is in comoving units, and we use $H(z_i)$ to
    translate it to velocity units.

  \item The emulator also returns an uncertainty associated to the prediction,
    that we add to the data covariance (after translating the emulator
    covariance to velocity units). 

  \item We use these ingredients to compute a Gaussian likelihood, and
    multiply it by the prior probability described above. 

\end{itemize}

We use \texttt{emcee} \citep{emcee} to run Monte Carlo Markov Chains,
and we use \texttt{GetDist} \citep{GetDist} to make contours plots with
marginalised posteriors.
\section{Cosmological information in the \lya\ \pd} 
\label{sec:direct}

In this section we follow the methodology described in Section \ref{sec:methods}
to fit cosmological parameters from a synthetic measurement of \pd.
We refer to these as \textit{direct fits}.

\begin{figure*}
  \centering
  \includegraphics[scale=0.6]{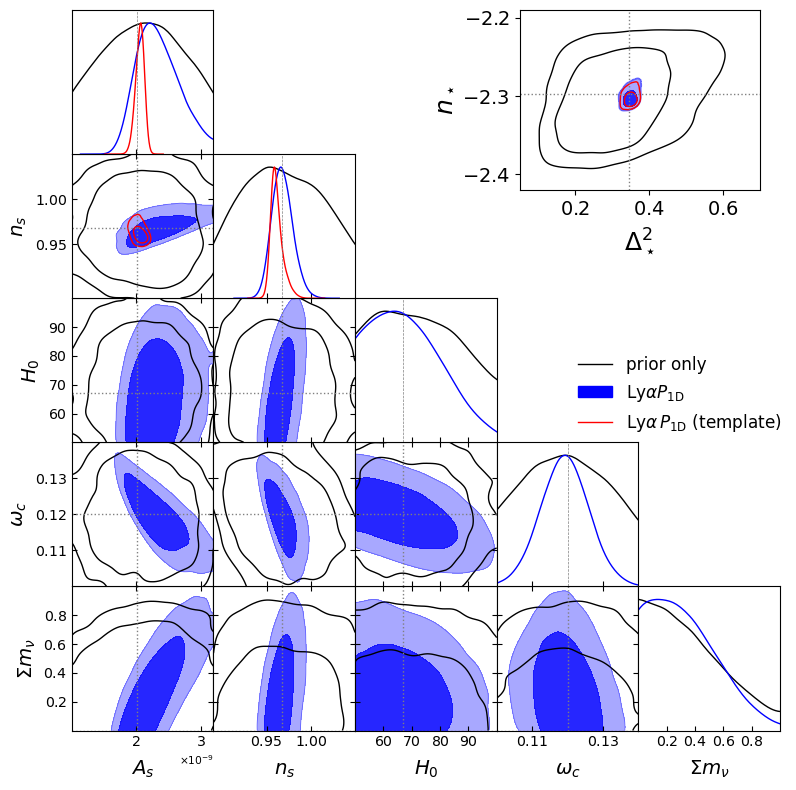}
  \caption{Direct fits to cosmological parameters from a mock \pd\ measurement
    from the \textit{Central} simulation. 
    We show the marginal posteriors on the five
    cosmological parameters that are being sampled. The top-right panel
    shows the marginal posteriors on the two derived parameters that will be
    used to compress the likelihood.
    Black lines correspond to running the analysis with only the prior,
    and dotted gray lines show the true values used to generate the mock.
    The blue contours show constraints from the \pd\ with all
    five cosmology parameters free. In the red contours, we show results
    where we use a \textit{template} cosmology, and fix all cosmology parameters to the values in the \textit{Central} simulation, except $A_s$ and $n_s$ which are kept free.
    We investigate the dependence of our posteriors on this choice of template in Figure \ref{fig:fid_cosmo}.
    For concision, we omit contours for the IGM parameters.
  }  
  \label{fig:p1d_vs_prior}
\end{figure*}

In Figure \ref{fig:p1d_vs_prior} we show the marginal constraints on cosmological
parameters when analysing mock data from the \textit{Central} simulation. 
In the standard analysis (blue), we vary five cosmological parameters
and eight nuisance parameters describing the IGM that are not shown.
For comparison, the black lines show the constraints from the priors
described in Table \ref{tab:priors}.

It is clear that \lya\ \pd\ alone cannot measure well these five cosmological
parameters, and that the results strongly depend on the choice of priors 
(the impact of the prior choice is discussed in Appendix \ref{app:priors}).
For instance, the constraints on $A_s$ are affected by the maximum value allowed
by the prior, and its lower bound is a consequence of the prior on neutrino
masses \sigmanu\ being positive.

On the top right corner of Figure \ref{fig:p1d_vs_prior} we also show the marginal
posteriors for the two derived parameters 
describing the linear power spectrum at $z=3$
(Equations \ref{eq:linP1} and \ref{eq:linP2}).
It is clear that adding \pd\ reduces dramatically the area of the prior
contours.
In the next sections we will refer to these as the \textit{compressed}
parameters, since they are able to compress most of the cosmological
information contained in the \lya\ \pd.

\subsection{Fixed template and fiducial cosmology}

The red contours in Figure \ref{fig:p1d_vs_prior} show a simplified version of the
analysis where only the primordial power parameters ($A_s$, $n_s$) and the
eight IGM parameters are varied.
In other words, we use a fixed \textit{template}
\footnote{This term is commonly used in redshift-space distortion (RSD)
analyses of galaxies to refer to analyses with fixed transfer functions
\citep{eBOSS2021}.}
for the linear power $P_L(z,k)$ and rescale it with these two parameters.
This analysis is significantly faster that the standard analysis, since we only
need to call CAMB a single time to compute the transfer function for the 
fiducial cosmology.

\begin{figure*}
  \centering
  \includegraphics[scale=0.45]{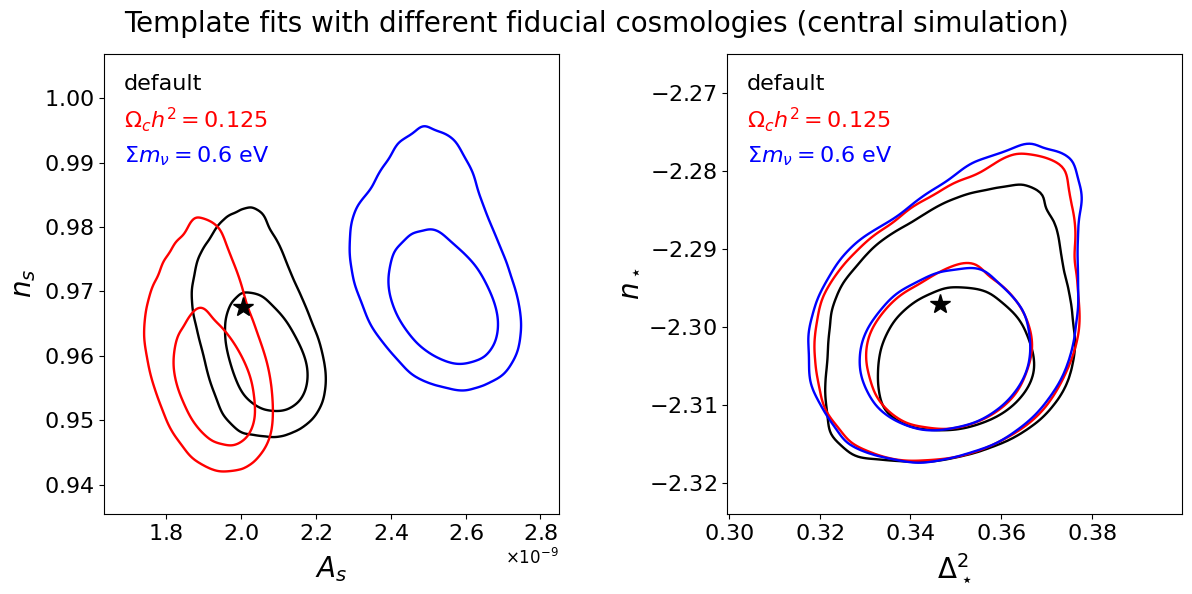}
  \includegraphics[scale=0.45]{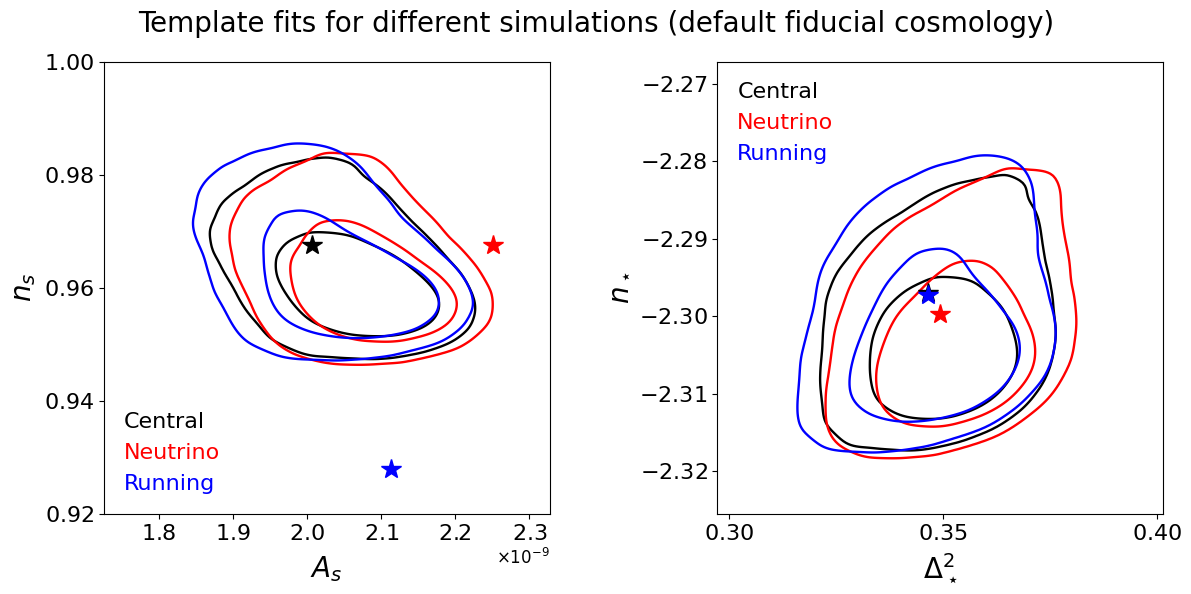}
  \caption{Top panels show marginal constraints on the primordial power
    parameters (left) and on the compressed parameters (right), when analysing
    the \textit{Central} simulation with different fiducial cosmologies.
    The fiducial cosmology in the \textit{default} analysis is the same one
    that was used to run the \textit{Central}, and stars mark the true value
    used in the simulation.
    When using a different fiducial cosmology, with an incorrect value of the
    CDM density (red) or neutrino masses (blue) we get biased constraints on
    primordial power parameters.
    On the other hand, the constraints on the compressed parameters are much
    less affected by the choice of fiducial cosmology.
    The bottom panels show equivalent constraints for the three test
    simulations, when analysed with the \textit{Central} cosmology as fiducial.
    Note that the \textit{Central} (black) and \textit{Running} (blue)
    simulations have the same values for the compressed parameters, but very
    different values for the primordial power spectrum (including different
    value of the running $\alpha_s$).
    }
  \label{fig:fid_cosmo}
\end{figure*}

The \textit{template} analysis can be seen as an analysis with infinitely 
tight priors on the other cosmological parmaters ($\omega_c$, $H_0$, \sigmanu).
While the constraints on the traditional cosmological parameters
($A_s$, $n_s$) are strongly affected by this change in the priors,
the constraints on the compressed parameters ($\Delta^2_\star$, $n_\star$, top
right panel) remain the same.

In this particular realisation of the analysis, we have use used a template
computed with the same cosmology that was used to run the simulation.
Even in the standard analysis (blue contours in Figure \ref{fig:p1d_vs_prior}) we
had to assume a value for the baryon density ($\omega_b=0.022$).
We will use the term \textit{fiducial cosmology} to refer to the cosmological
parameters that are being kept fixed in the analysis.
Obviously in a real analysis the true cosmology is not
known, and so we next test the effect of changing this
fiducial comsmology on our results.

In the top panels of Figure \ref{fig:fid_cosmo} we redo the \textit{template}
analysis when using different fiducial cosmologies, with the wrong CDM density
(in red) or the wrong sum of the neutrino masses (in blue).
While there is a clear bias on the primordial power parameters (left), the
compressed parameters are much less affected by the choice of fiducial
cosmology.

The bottom panels of the same figure show a \textit{template} analysis for 
the three test simulations described in Table \ref{tab:sims}.
In all three analyses we use the \textit{Central} cosmology as our fiducial
cosmology.
As can be seen in the left bottom panel, this results in biased posteriors for
the primordial power parameters in the \textit{Neutrino} and \textit{Running}
simulations (stars identify the true values used in each simulation).
However, the marginal posteriors of the compressed parameters are again
recovered successfully (right bottom panel).
These marginal posteriors in the bottom right panel will be used in the next
section.
\section{Joint analysis with CMB}
\label{sec:cmb}

In the previous sections we discussed cosmological fits from the \lya\ \pd\
alone, with only weak priors on cosmological parameters.
We showed that we can measure very well the amplitude ($\Delta^2_\star$) and
slope ($n_\star$) of the linear power spectrum around $z_\star=3$ and
$k_p=0.009\ikms$, and that the constraints on these \textit{compressed}
parameters were unbiased, and do not depend on our choice of priors or 
fiducial cosmology.

In this section we discuss joint cosmological analysis with anisotropies in
the Cosmic Microwave Background (CMB).
CMB and \pd\ measurements are very complementary, since together they cover
a very wide range of scales and redshifts.
This has made these joint analyses very popular in the past 
\citep{Phillips2001,Verde2003,Spergel2003,Seljak2005,Seljak2006,PD2015,PD2015b,PD2020}, and they are forecasted to
provide some of the tightest constraints on the sum of the neutrino masses 
and on the running of the spectral index from future surveys
\citep{2014JCAP...05..023F}.

\begin{figure*}
  \centering
  \includegraphics[scale=0.55]{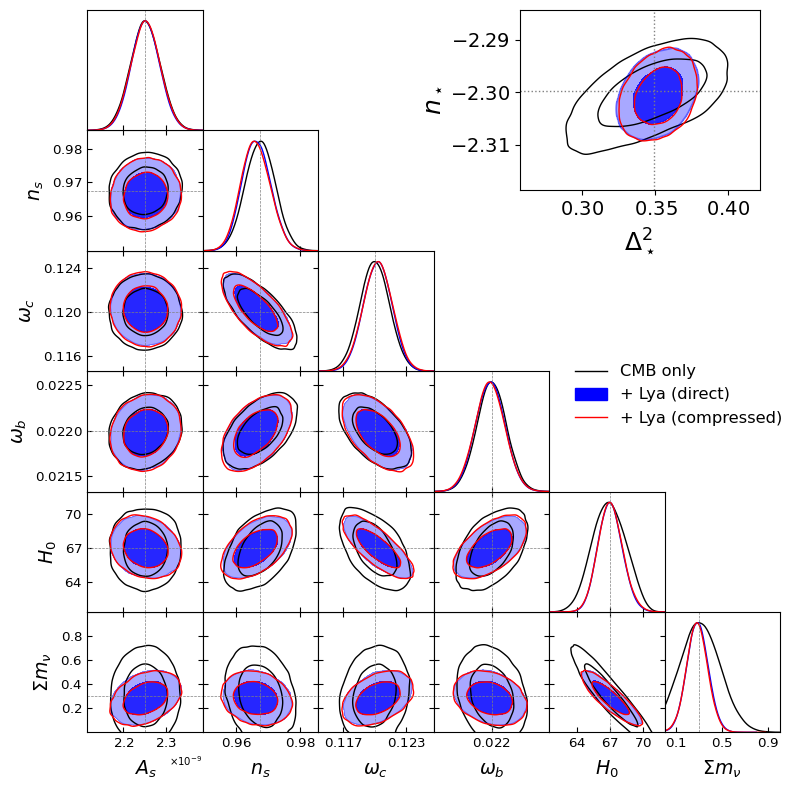}
  \caption{Cosmological constraints from CMB + \lya\ \pd, for mock data
  from the \textit{Neutrino} simulation. 
  Blue contours use a direct \pd\ likelihood, while red contours use the
  marginal posterior on linear power parameters ($\Delta^2_\star$,$n_\star$).
  Black contours show the CMB-only results, with the grey dashed lines
  representing the values in the mock simulation.
  The primordial power is assumed to have no running in this analysis.}
  \label{fig:cmb_mnu}
\end{figure*}

\begin{figure*}
  \centering
  \includegraphics[scale=0.55]{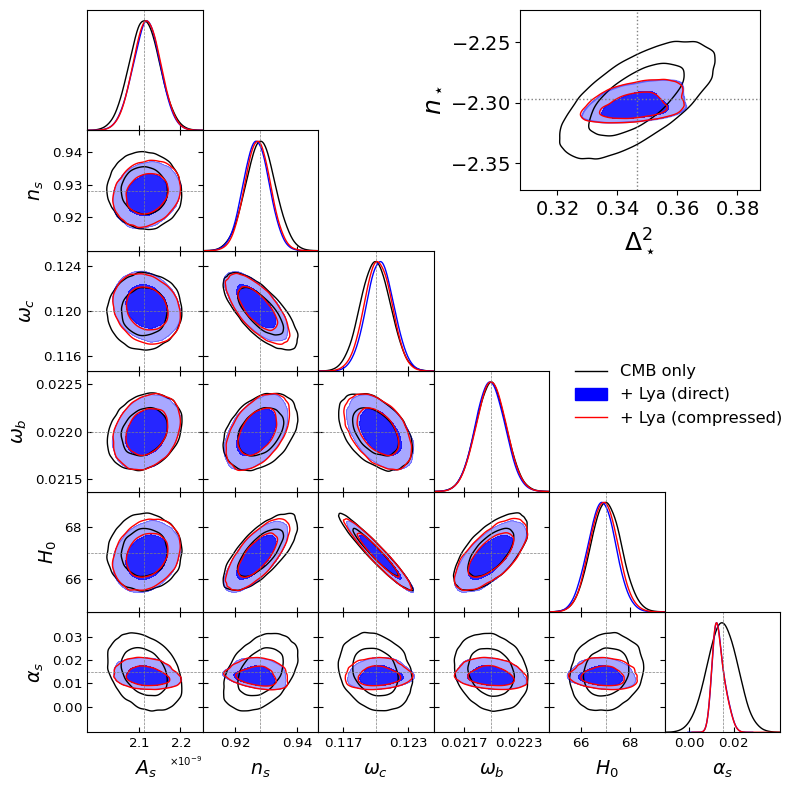}
  \caption{Same as Figure \ref{fig:cmb_mnu}, except now using the mock data from the 
  \textit{Running} simulation.
  Neutrino masses are fixed to $0$ in this analysis.}
  \label{fig:cmb_nrun}
\end{figure*}

Instead of using an actual CMB likelihood, for simplicity we use a Gaussian
likelihood on the relevant cosmological parameters.
The Gaussian likelihood uses a covariance matrix obtained from the
official \textit{Planck} chains
\footnote{For chains with massive neutrinos, we have computed the covariance
around its best-fit value ($\summnu=0$) and not around its mean.}.
The centre of the Gaussian has been set to the values used in the different
test simulations described in Section \ref{sec:methods}.
The approximated CMB likelihood can be seen in solid black contours in Figures
\ref{fig:cmb_mnu} (free neutrino mass) and \ref{fig:cmb_nrun} (free running).

The results in Figure \ref{fig:cmb_mnu} are from a joint analysis
of the CMB and our mock \pd\ from the \textit{Neutrino} simulation, when
varying 6 cosmological parameters ($A_s$, $n_s$, $\omega_b=\Omega_b h^2$, 
$\omega_c=\Omega_c h^2$, \sigmanu and $\thetamc$), with the priors described
in Table \ref{tab:priors}.
Even though we sample $\thetamc$, we plot the contours for $H_0$, computed
as a derived parameter.

The blue contours show a joint fit using the direct \pd\ likelihood, i.e., 
we have varied at the same time the cosmological parameters and the
8 nuisance parameters that were also used in Section \ref{sec:direct} to 
describe the uncertainties in the physics of the IGM.

The red contours, on the other hand, use the marginal posterior on the linear
power parameters ($\Delta^2_\star$, $n_\star$) obtained from the \lya\ \pd\
alone. 
In more detail, to obtain the red contours we:
\begin{itemize}
  \item Run a \textit{template} fit to the \lya\ \pd\ alone, varying 8 IGM
    parameters and 2 cosmological parameters ($A_s$, $n_s$), as described
    in Section \ref{sec:direct}.
  \item Use a Kernel Density Estimator (KDE, from \texttt{SciPy} \citep{SciPy})
    to model the marginal posteriors on the two compressed parameters
    ($\Delta^2_\star$, $n_\star$), shown in the bottom right panel of Figure
    \ref{fig:fid_cosmo}.
  \item Run a joint analysis of CMB and the marginal \lya\ \pd\ posterior,
    varying 6 cosmological parameters ($A_s$, $n_s$, $\omega_b$,
    $\omega_c$, \sigmanu\ and $\thetamc$).
    It is important to note that in this last step one does not need to use
    an emulator, or worry about the nuisance parameters describing the IGM;
    these have already been marginalised over in the previous steps.
\end{itemize}

It is remarkable that in both cases we recover the true value for the sum 
of the neutrino masses, even though our emulator was constructed from simulations
that assume massless neutrinos. 
It is also remarkable how similar are the joint constraints when using 
the \textit{direct} (blue) and \textit{compressed} (red) likelihoods.
This implies that there is negliglble loss of cosmological information
when compressing the \pd\ into marginal constraints on the linear power spectrum.

In Figure \ref{fig:cmb_nrun} we present a similar analysis for the \textit{Running}
simulation, where we have assumed that neutrinos are massless but we have
explored models with running of the spectral index $\alpha_s$.
Here again we recover the right cosmology, and both approaches give very
consistent results.

The effect on posteriors of including \lyaf\ information is slightly different in Figures \ref{fig:cmb_mnu} and \ref{fig:cmb_nrun}. Whilst we do not include a figure for this, in the case of a simple flat $\Lambda$CDM model, the constraints from the CMB alone on $\Delta^2_\star$ and $n_\star$ are already very good, and the \lyaf\ does not provide a significant improvement. It is in the analysis of extended models, such as the two we consider in this work, that the contribution from \lyaf\ becomes important. In the case of free neutrino mass, the improvement is mostly in $\Delta_\star^2$. This is because free neutrino mass opens up a degeneracy in the amplitude of the late-time power spectrum obtained from CMB-only analysis, and including \lyaf\ information breaks this degeneracy. For the case of free $\alpha_s$, the information from a smaller pivot scale is essential in breaking the degeneracy between $\alpha_s$ and $n_\star$.
\section{Discussion} 
\label{sec:discussion}

In Section \ref{sec:direct} we have shown that the \lya\ \pd\ can robustly measure
two parameters describing the amplitude and slope of the linear power spectrum
at a central redshift $z_\star=3$, and around a pivot point
$k_\star=0.009 \, \ikms$ defined in velocity units.
We have shown that we recover unbiased results independent of the fiducial
cosmology assumed in the fits, even when analysing models that were not
included in the training of our \texttt{LaCE} emulator.

In Section \ref{sec:cmb} we have shown that, in the context of joint analyses with
CMB data, the cosmological information in the \lya\ \pd\ can be captured with
the marginalised posteriors of these two parameters.
We have explicitly shown that this is the case for the two single-parameter
extensions to the $\Lambda$CDM model where the \pd\ is forecasted to contribute
the most \citep{2014JCAP...05..023F}: 
models with massive neutrinos (Figure \ref{fig:cmb_mnu}) and models with running of
the spectral index of primordial fluctuations (Figure \ref{fig:cmb_nrun}).

The compression is successful as we are able to approximate the expansion
and growth rates in the $2<z<5$ regime using a fixed fiducial model, due to the
fact that the Universe is close to Einstein de-Sitter in this regime.
At significantly higher data precision, one would expect this approximation to
break down, and therefore the compression to fail. In this case, including the
extended parameters discussed in \ref{app:extended} might capture the missing information.
However, the data covariance we use in this work will not be surpassed by any current
or proposed experiment, so we leave quantifying the regime in which the compression
fails to a future work.

Exotic cosmological models might require more complex implementations
of the emulation and compression schemes discussed in this work.
For instance, models with either warm or fuzzy dark matter predict
that the linear power spectrum could be strongly suppressed on sub-megaparsec
scales, and the \lyaf\ has provided some of the tightest constraints on these
models \citep{Viel2013,Irsic2017b,Irsic2017c,Murgia2018,PD2020,Rogers2021b}.
In order to use the \texttt{LaCE} emulator in these studies, one would need
to add extra emulator parameters describing the suppression of the linear power,
and run extra simulations exploring them.
Equivalently to $\Delta^2_\star$ and $n_\star$, one would need to define other
\textit{compressed} parameters to capture the relevant information present in
the \pd\ likelihood.
Since the \pd\ measurements are naturally carried out in velocity units,
these extra parameters would also need to be defined in velocity units,
otherwise the cutoff scale would depend on the assumed model of the expansion
rate $H(z)$.

In the next few years, the Dark Energy Spectroscopic Instrument (DESI) will
measure with unprecedented accuracy the \lya\ \pd, enabling very precise
constraints on the linear power spectrum of matter fluctuations around $z=3$.
We expect that the compression scheme discussed here will significantly increase
the impact of these measurements, and it will simplify joint analyses with
external datasets.

\begin{acknowledgments}
{The authors thank Patrick McDonald, Pablo Lemos Portela, and the \lya\ working group of DESI
for useful discussions.
CP acknowledges support by NASA ROSES grant 12-EUCLID12-0004.
AFR acknowledges support from the Spanish Ministry of Science and Innovation
through the program Ramon y Cajal (RYC-2018-025210)
and from the European Union’s Horizon Europe research and 
innovation programme (COSMO-LYA, grant agreement 101044612).
IFAE is partially funded by the CERCA program of the Generalitat de Catalunya.
This manuscript has been co-authored by Fermi Research Alliance, LLC
under Contract No. DE-AC02-07CH11359 with the U.S. Department of Energy,
Office of Science, Office of High Energy Physics.
This work was partially supported by the Visiting Scholars Award Program of the Universities Research Association, and by funding from the UCL
Cosmoparticle Initiative.
This work used computing facilities
provided by the UCL Cosmoparticle Initiative.
The simulations were run using the Cambridge Service for Data Driven Discovery
(CSD3), part of which is operated by the University of Cambridge Research
Computing on behalf of the STFC DiRAC HPC Facility (\url{www.dirac.ac.uk}).
The DiRAC component of CSD3 was funded by BEIS capital funding via STFC capital
grants ST/P002307/1 and ST/R002452/1 and STFC operations grant ST/R00689X/1.
DiRAC is part of the National e-Infrastructure.}\end{acknowledgments}

\appendix
\section{Impact of prior choice}
\label{app:priors}

\begin{figure}
  \centering
  \includegraphics[scale=0.4]{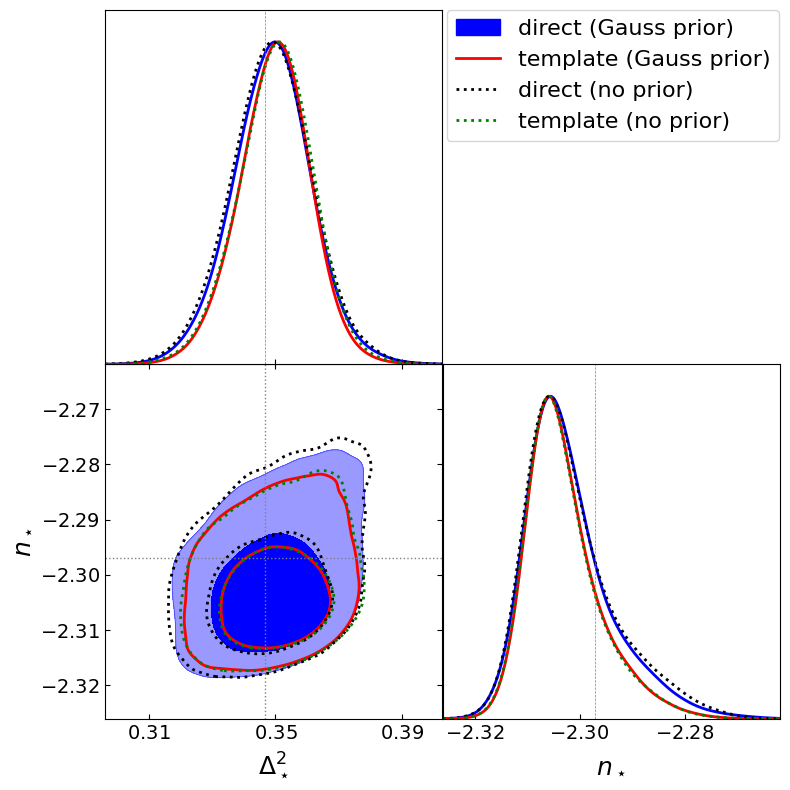}
  \caption{Marginalised 1D and 2D posterior distributions on compressed
    parameters, corresponding to analyses of the \textit{Central} mock data.
    In blue we show the constraints from a direct \pd\ analysis using the
    loose Gaussian priors, and in red we show the constraints from an
    equivalent template fit (fixed values for $\omega_c$, $H_0$ and \sigmanu);
    these 2D contours were already presented in the top right panel of 
    Figure \ref{fig:p1d_vs_prior}.
    The black (green) dotted contours show the constraints from a direct
    (template) fit when not using any Gaussian prior, and demonstrate that
    the role of the Gaussian prior on the compressed constraints is very minor.}
    \label{fig:priors}
\end{figure}

The results presented in the main text used a Gaussian prior described
in Table \ref{tab:priors}. 
In Figure \ref{fig:priors} we demonstrate that the marginalised posteriors for the
compressed parameters are not affected by this prior.

We start by showing the results from a direct analysis (blue contours) and
a template analysis (red contours) that include the Gaussian prior;
these are the contours already presented in the top right panel of 
Figure \ref{fig:p1d_vs_prior}.
These can be compared respectively to the black and green dotted contours,
where we have not included the Gaussian prior.
\section{Impact of cosmic variance in the emulator predictions}
\label{app:seed}

\begin{figure}
  \centering
  \includegraphics[scale=0.4]{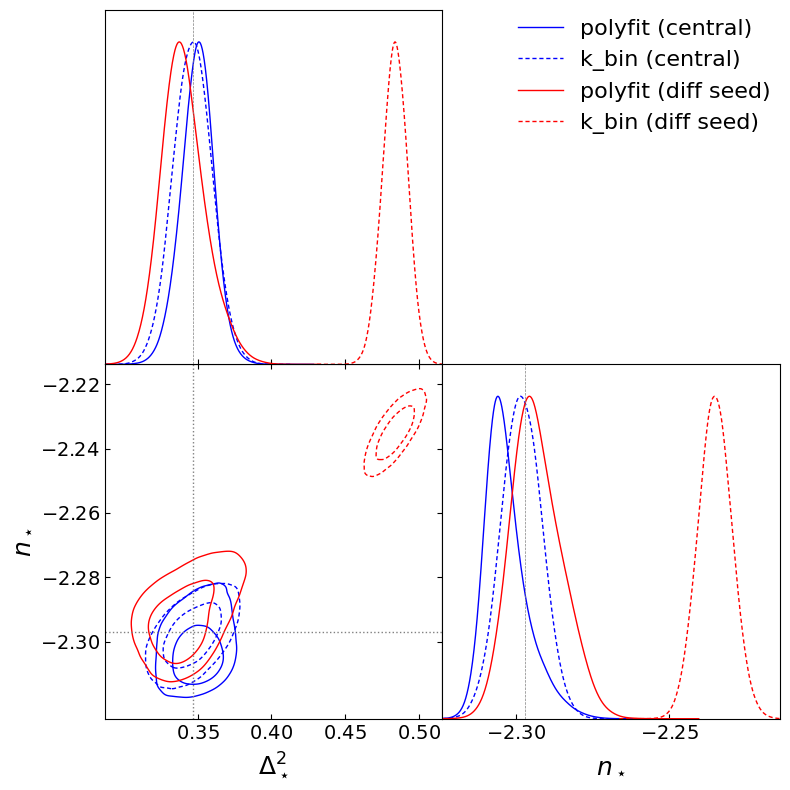}
  \caption{Marginalised 1D and 2D posterior distributions on compressed
    parameters, corresponding to template fits to the \textit{Central} mock
    data (in blue) discussed in Figure \ref{fig:p1d_vs_prior},
    and fits to mock data with different random phases (diff seed, in red).
    In solid lines show the constraints when using the \textit{polyfit}
    framework used in the main text, where we emulate the coefficients of
    polynomial fits to \pd.
    The dotted lines, on the other hand, use the \textit{k\_bin}
    framework that was used in \citet{Pedersen2021}, where we emulate the value
    of \pd\ on a grid of wavenumbers.
    While both frameworks give consistent results when analysing the
    \textit{Central} simulation, it is clear that the \textit{k\_bin} emulator
    gives biased results when analysing simulations with different random
    phases (dotted red contours).}
    \label{fig:seed}
\end{figure}

In the main text we have analysed simulations that had initial conditions
generated with the same random phases than the simulations used to train the
\texttt{LaCE} emulator.
In order to study the impact of cosmic variance in the emulator predictions, 
in \ref{fig:seed} we show the results when analysing a test simulation 
\textit{diff seed} (red contours) that has the same physics than the 
\textit{Central} simulation (blue contours), but has different random phases
in the initial conditions. 

In the same figure we also compare the results when using two different
implementations of the \texttt{LaCE} emulator:
the \textit{polyfit} framework (solid lines), used in the main text, emulates
the value of the coefficients of polynomial fits describing the \lya\ \pd
(Equation \ref{eq:polyfit});
the \textit{k\_bin} framework (dotted lines), used in \citet{Pedersen2021},
directly emulates the value of \lya\ \pd\ on a fine grid of wavenumbers.

It is clear that the \textit{k\_bin} emulator gives biased results, probably
because it is trying to fit different noise spikes than the ones used in the
training sample.
On the other hand, the \textit{polyfit} emulator is able to give unbiased
results even when analysing mock data with different cosmic variance.

\section{Extended compression schemes}
\label{app:extended}

In Section \ref{sec:direct} we have proposed to compress the cosmological information
in \pd\ into two parameters describing the amplitude ($\Delta^2_\star$) and
slope ($n_\star$) of the linear power spectrum at $z_\star=3$,
around a pivot point $k_\star=0.009 ~\ikms$.
We have shown in Section \ref{sec:cmb} that this compression is lossless in the
context of joint analyses with the CMB with free neutrino masses (\sigmanu), 
or free running of the primordial power spectrum ($\alpha_s$).
In Section \ref{sec:discussion} we mentioned that one might need to add extra
parameters describing the shape of the linear power at $z_\star$.
For instance, a third parameter describing the curvature around the pivot
point \citep{McDonald2005}, or a cut-off to describe the small-scales
suppression in non-cold dark matter models.
In this appendix, instead, we discuss possible extensions to capture other
cosmological information beyond the shape of the linear power at $z_\star$.

Measurements of the \lya\ \pd\ typically cover a wide range of redshifts.
For instance, \citet{Chabanier2019} measured \pd\ from $z=2.2$ to $z=4.6$.
It might seem surprising that we can capture all the cosmological information
when parameterising the linear power spectrum at a single redshift $z_\star=3$.
Moreover, while the shape of the linear power spectrum is constant when
described in comoving units, the same is not true when described in velocity
units.
Our pivot scale $k_\star$ correspond to different comoving separations at
different redshifts, and one could imagine measuring $H(z)/(1+z)$ from
the redshift evolution of the shape of the linear power in velocity units.

In order to capture information from these two effects, we introduce two
extra parameters.
We parameterise the growth of structure around $z_\star$ with the logarithmic
growth rate $f_\star=f(z_\star)$, defined as usual:
\begin{equation}
  f(z) = \frac{\partial \ln D(z)}{\partial \ln a(z)} ~,
  \label{eq:fstar}
\end{equation}
with $f_\star=1$ in an Einstein-de Sitter (EdS) universe.

Similarly, we parameterise the evolution of the expansion rate around $z_\star$
in terms of $g_\star=g(z_\star)$, defined as:
\begin{equation}
  g(z) = \frac{\partial \ln H(z)}{\partial \ln (1+z)^{3/2}} ~,
  \label{eq:gstar}
\end{equation}
such that $g_\star=1$ corresponds again to an EdS universe.

\subsection{Template fits with $f_\star$ and $g_\star$}

Instead of looking at posteriors of $f_\star$ and $g_\star$ computed 
as derived parameters in fits for a particular model, we would like to
directly sample these without assuming any cosmological model.
For instance, in a $\Lambda$CDM universe, without curvature or massive
neutrinos, both $f_\star$ and $g_\star$ would be just a function of $\Omega_m$.
However, more exotic models could decouple the linear growth from the
expansion of the universe, making these parameters independent.

Therefore, in this appendix we directly sample the four compressed parameters
($\Delta^2_\star$, $n_\star$, $f_\star$, $g_\star$) and the same eight nuisance
parameters used in the main text to model the IGM.
We use a uniform prior range of $[0.24,0.47]$, $[-2.352,-2.25]$, $[0.9,1.0]$,
$[0.9,1.0]$ for each parameter respectively.
The details of how we do this are detailed later in Section \ref{ss:recons}.

\begin{figure}
  \centering
  \includegraphics[scale=0.5]{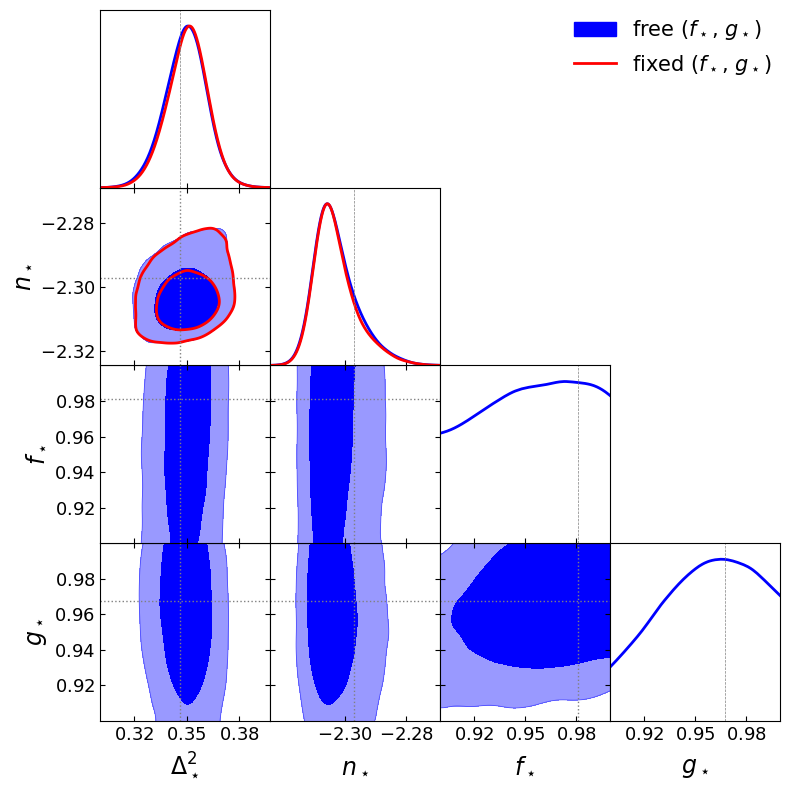}
  \caption{Marginalised 1D and 2D posterior distributions on compressed
    parameters when analysing the mock dataset from the \textit{Neutrino}
    simulation.
    In the red contours we fix $f_\star$ and $g_\star$ to the fiducial values
    described in the text.
    The dashed lines show the true values in the mock simulation, and the
    shaded areas of the 1D posteriors show the $68\%$ credible region.}
    \label{fig:extended_nu}
\end{figure}

In figure \ref{fig:extended_nu} we show constraints on compressed parameters,
after marginalising over the IGM. 
We use as a mock dataset the \textit{Neutrino} simulation, and show two sets
of constraints.
In red, we have fixed $f_\star$ and $g_\star$ to the values of the fiducial
cosmology ($f_\star=0.981$, $g_\star=0.968$), whereas in blue they are
left as free parameters. 
The dashed lines show the true values in the mock simulation 
($f_\star=0.989$, $g_\star=0.969$).
We note that $f_\star$ is very poorly constrained, implying that the \pd\
alone is not highly sensitive to the redshift evolution of the linear power
spectrum.
This result is consistent with the findings of \citet{McDonald2005}, although
we confirm that this is still the case when using high precision datasets.
The posterior for $g_\star$ is slightly better constrained, although it can
only rule out very low values of $g_\star < 0.9$.
Additionally there is very little effect on the posteriors for $\Delta^2_\star$
and $n_\star$ when marginalising over $f_\star$ and $g_\star$ when compared to
fixing them.

Note that the red contours were constructed assuming the wrong background
cosmology (wrong values of $f_\star$ and $g_\star$), but that the constraints
on $\Delta^2_\star$ and $n_\star$ are nevertheless unbiased.

\subsection{Reconstructing the linear power spectrum}
\label{ss:recons}

Here we describe the procedure for mapping from a set of values for
the compressed parameters ($\Delta^2_\star$, $n_\star$, $f_\star$, $g_\star$)
to the 11 pairs of emulator parameters ($\Delta^2_p$, $n_p$) values required
to generate theoretical predictions for the \pd\ from the emulator, one at
each redshift.
This is done using a fiducial cosmology, as outlined below.
We will use $k$ to refer to the (modulus of the) 3D wavenumbers in comoving
coordinates, i.e., in \Mpc. 
We will use $q$ to refer to the same wavenumber in velocity units. 
They are related by:
\begin{equation}
 k = \frac{H(z)}{1+z} ~ q = M(z) ~ q ~.
\end{equation}
$M(z)$ will play an important role in this discussion.

We will use $P(k)$ to refer to (3D) power spectra in comoving units, i.e.,
with units of Mpc$^3$.
We will use $Q(q)$ to refer to (3D) power spectra in velocity units, i.e.,
with units of (kms$^{-1}$)$^3$.
They are related by:
\begin{equation}
 Q(z,q) = M^3(z) ~ P(z,k=M(z)q) ~.
\end{equation}

In our code we will use a fiducial cosmology as a reference, and parameterise
our models as deviations from that cosmology.
We will use either subscripts $_0$ or superscripts $^0$ to identify functions
for the fiducial cosmology.
We address changes to the shape of the power spectrum, as described by
$\Delta^2_\star$, $n_\star$ (and $\alpha_\star$) first, and then later
address changes to the redshift evolution using $f_\star$ and $g_\star$.
We can now define the ratio of the linear power between any model and the
fiducial one, at the central redshift $z_\star$, and in velocity units:
\begin{equation}
 B(q) = \frac{Q_\star(q)}{Q^0_\star(q)} ~.
\end{equation}
This will be another important function, tightly related to the linear
power parameters that we will end up using.

We fit a second order polynomial to the logarithm of the linear power
spectrum at $z_\star$, in velocity units, around a pivot point $q_\star$.
By default we use $z_\star=3$ and $q_\star=0.009$ s/km, and we fit the polynomial
in a range of wavenumbers defined as $q_\star / 2 < q < 2 q_\star$
\footnote{We do the fit using \texttt{numpy.polyfit}}.

\begin{equation}
 Q_\star(q) \approx A \left( \frac{q}{q_\star} \right)^{n_\star + \alpha_\star /2 \ln (q/q_\star)} ~,
\end{equation}
or equivalently
\begin{equation}
 \ln Q_\star(q) \approx \ln A + \left[ n_\star + \alpha_\star /2 \ln (q/q_\star) \right] \ln (q/q_\star)~.
\end{equation}
$n_\star$ is the first log-derivative around $q_\star$, and $\alpha_\star$
is the second log-derivative around the same point.
Note that the polynomial fit, however, returns
($\ln A$, $n_\star$, $\alpha_\star /2$).
Finally, we define a dimensionless parameter describing the amplitude, 
$\Delta^2_\star = A ~ q_\star^3 / (2\pi^2)$. When reconstructing the linear power spectrum using a fiducial cosmology,
we use differences in the shape parameters with respect to the fiducial ones:
\begin{align}
 \ln B(q) & = \ln Q_\star(q) - \ln Q^0_\star(q)       \nonumber \\
  & \approx \left( \Delta^2_\star - \Delta^2_{\star~0} \right)
    + \left[ \left( n_\star - n^0_\star \right) 
    + \frac{\alpha_\star - \alpha^0_\star}{2} \ln (q/q_\star) \right] \ln (q/q_\star)~.
\end{align}

We are also concerned with reconstructing the linear power spectrum at redshifts other than $z_\star$.
We ignore neutrinos for now, and work with just the CDM+baryon power spectrum.
In this case we can use the linear growth factor $D(z)$, defined as
\begin{equation}
 P(z,k) = \left[\frac{D(z)}{D_\star}\right]^2 P_\star(k) ~,
\end{equation}
where in general functions $y_\star = y(z_\star)$. We can write the power spectrum at an arbitrary redshift as  a function of the fiducial one:
\begin{equation}
\begin{split}
    Q(z,q) &= M^3(z) ~ P(z,k=M(z)q)    \\
     &= M^3(z) ~ \left[\frac{D(z)}{D_\star}\right]^2 P_\star(k=M(z)q) \nonumber \\
     &= \left[\frac{M(z)}{M_\star}\right]^3 \left[\frac{D(z)}{D_\star}\right]^2 
       Q_\star(q^\prime=M(z)/M_\star q)    \nonumber \\
     &= \left[\frac{M(z)}{M_\star}\right]^3 \left[\frac{D(z)}{D_\star}\right]^2 
       B(q^\prime=M(z)/M_\star q) ~ Q^0_\star(q^\prime=M(z)/M_\star q)  \nonumber \\
     &= \left[\frac{M(z)}{M_\star}\right]^3 \left[\frac{D(z)}{D_\star}\right]^2 
       B(q^\prime=M(z)/M_\star q) \left[ M_\star^0 \right]^3 
       P^0_\star(k=M_\star^0 M(z) / M_\star q)  \nonumber \\
     &= \left[\frac{M(z)}{M_\star}\right]^3 \left[\frac{D(z)}{D_\star}\right]^2 
       B(q^\prime=M(z)/M_\star q) \left[ M_\star^0 \right]^3 
       \left[\frac{D^0_\star}{D_0(z)}\right]^2 
       P^0(z,k=M_\star^0 M(z) / M_\star q)  \nonumber \\
     &= \left[\frac{M(z)}{M_\star}\right]^3 \left[\frac{D(z)}{D_\star}\right]^2 
       B(q^\prime=M(z)/M_\star q) \left[ \frac{M_\star^0}{M_0(z)} \right]^3 
       \left[\frac{D^0_\star}{D_0(z)}\right]^2 
       Q^0(z,q^\prime= (M_\star^0 M(z)) / (M_\star M_0(z)) q)  \nonumber \\
     &= \left[m(z) \right]^3 \left[d(z)\right]^2 
       B(q^\prime=m(z) M_0(z)/M^0_\star q) ~ Q^0(z,q^\prime=m(z)q) \nonumber ~,
\end{split}
\end{equation}
where for convenience we have defined two functions,
\begin{equation}
 m(z) = \frac{M(z)}{M_\star} \frac{M_\star^0}{M_0(z)} 
\end{equation}
and 
\begin{equation}
 d(z) = \frac{D(z)}{D_\star} \frac{D^0_\star}{D_0(z)} ~,
\end{equation}
that describe differences in expansion rate and in linear growth respectively.

Using the definition of $g_\star$ in Equation \ref{eq:gstar}, we approximate $m(z)$
using the difference of $g_\star$ between the input and the fiducial cosmology
as:
\begin{equation}
 \ln m(z) \approx \frac{3}{2} \left( g_\star - g^0_\star \right) 
    \ln \left( \frac{1+z}{1+z_\star} \right) ~,
\end{equation}
or equivalently
\begin{equation}
 m(z) \approx \left( \frac{1+z}{1+z_\star} \right)^{3/2 ( g_\star - g^0_\star)} ~.
\end{equation}
Similarly, we approximate $d(z)$ using the difference of $f_\star$ between
the input and the fiducial cosmology as:
\begin{equation}
 \ln d(z) \approx - \left( f_\star - f^0_\star \right) 
    \ln \left( \frac{1+z}{1+z_\star} \right) ~,
\end{equation}
or equivalently
\begin{equation}
 d(z) \approx \left( \frac{1+z}{1+z_\star} \right)^{- ( f_\star - f^0_\star)} ~.
\end{equation}
With these equations, for a given set of ($\Delta^2_\star$, $n_\star$,
$\alpha_\star$, $f_\star$, $g_\star$), $Q(z,q)$ can be estimated.
We then use the approximation of $m(z)$ to convert the velocity unit
power spectrum to a comoving power spectrum, and fit a polynomial over the
range $k_p/2 < k < 2k$ to obtain values for $\Delta^2_p$ and $n_p$. 
Note that the emulator returns a \pd\ in comoving units.
The final step is to convert this into velocity units, once again using
the above approximation of $m(z)$.
This reconstruction process and the composite approximations have been
compared against the true values generated in \texttt{CAMB}, and we verified
that they are accurate to within the percent level across all redshifts and
extended model spaces considered in this paper.

\bibliography{main}{}
\bibliographystyle{aasjournal}

\end{document}